# Molecular Dynamics Simulation on a Glassforming NiZr-System: *Diffusion Coefficients and Critical Temperature*

A.B. Mutiara

Dept. of Informatics Engineering, Fac. of Industrial Technology, University of Gunadarma,
Jl. Margonda Raya No 100, Depok
Tel. 78881112 ext.309 email: amutiara@staff.gunadarma.ac.id

*Abstract*– **The diffusion coefficients of Ni $D_{Ni}$ and Zr $D_{Zr}$, and also the critical temperature $T_c$ of the system, as results of data analysis from Molecular Dynamics (MD) Simulation, are presented. An *NpT*-Ensemble of 648 atoms is simulated in a Box with $L \sim 26{,}2$ Å by using Stillinger-Weber potential model with adopting paramaters for NiZr-System that based on results of "first-principle" electron teori from Hausleitner and Hafner. The present simulations are carried out by numerically integrating the Newtonian equation of atoms, and for integrating a fifth order predictor-corrector algorithm is used with time step $\Delta t = 2.5 \, 10^{-15}$ s. Then diffusion coefficients and a critical temperature $T_c$ are analyzed qualitatively on the basis of mode-coupling theory (MCT) for glass transition (GT). One of MCT-predictions is that temperature dependence of Diffusion coefficients follows a "power-law" with non-universal exponent parameter γ, namely $D_i \sim |T-T_c|^\gamma$. From analysis results it is found that $T_c \sim 950$ K with γ ~ 1.8 for Ni-atom and γ ~ 2.0 for Zr-atom.**

## I. INTRODUCTION

The phenomena behind the liquid-glass transition and the nature of the glassy state are not fully understood, despite the progresses of recent years [1]. In contrast to usual phase transformations the glass transition seems to be primarily dynamical in origin and therefore new theoretical approaches have to be developed for its description [2,3]. Several theoretical models have been proposed to explain the transition and the corresponding experimental data. The latter concern both *the temperature dependence of particular properties*, such as the shear viscosity and the structural relaxation time, and the *time dependent response* as visible in the dielectric susceptibility, inelastic neutron scattering, and light scattering spectra investigations. The spectral measurements have been extended to cover the large frequency range from below the primary α-relaxation peak up to the high-frequency region of microscopic dynamics dominated by vibrational modes [4,5].

One of the promising theoretical approaches in this field is the mode coupling theory (MCT). The MCT originally was developed to model critical phenomena [6,7]. The non-classical behavior of the transport properties near the critical point was thought to be caused by nonlinear couplings between slow (hydrodynamic and order parameter) modes of the system. In later years, the MCT was found to be applicable more generally to describe nonlinear effects in dense liquids [8] and nonhydrodynamic effects in the case of the glass transition.

In its simplest ('idealized') version, firstly analyzed in the schematic approach by Bengtzelius et al. [9] and independently by Leutheusser [10], the MCT predicts a transition from a high temperature liquid (``ergodic") state to a low temperature arrested (``non-ergodic") state at a critical temperature $T_c$. Including transversale currents as additional hydrodynamic variables, the full MCT shows no longer a sharp transition at $T_c$ but all structural correlations decay in a final α-process [11]. Similar effects are expected from inclusion of thermally activated matter transport, that means diffusion in the arrested state [12,13].

In the full MCT, the remainders of the transition and the value of $T_c$ have to be evaluated, e.g., from the approach of the undercooled melt towards the idealized arrested state, either by analyzing the time and temperature dependence in the β-regime of the structural fluctuation dynamics [14-16] or by evaluating the temperature dependence of the so-called $g_m$-parameter [17,18]. There are further posibilities to estimates $T_c$, e.g., from the temperature dependence of the diffusion coefficients or the relaxation time of the final α-decay in the melt, as these quantities for $T > T_c$ display a critical behaviour $|T-T_c|^{\pm\gamma}$. However, only crude estimates of $T_c$ can be obtained from these quantities, since near Tc the critical behaviour is masked by the effects of transversale currents and thermally activated matter transport, as mentioned above.

On the other hand, as emphasized and applied in [19-21], the value of $T_c$ predicted by the idealized MCT can be calculated once the partial structure factors of the system and their temperature dependence are sufficiently well known. Besides temperature and particle concentration, the partial structure factors are the only significant quantities which enter the equations of the so-called nonergodicity parameters of the system. The latter vanish identically for temperatures above $Tc$ and their calculation thus allows a rather precise determination of the critical temperature predicted by the idealized theory.

We here investigate a molecular dynamics (MD) simulation model adapted to the glass-forming $Ni_{20}Zr_{80}$ transition metal system. The $Ni_xZr_{1-x}$-system is well studied by experiments [22,23] and by MD-simulations [24-28], as it is a rather interesting system whose components are important constituents of a number of multi-component 'massive' metallic glasses. In the present contribution we consider, in particular, the x=0.2 composition and concentrate on the determination of Tc from evaluating and analyzing diffusion coefficients.

In the literature, the investigation of $Tc$-estimates already exist [19-21] for two systems (e.g. from MD-simulations for a soft spheres model [19] and, from MD-simulations for a binary Lennard-Jones system [21]. Regarding this, the present investigation is aimed at clarifying the situation for at least one of the important metallic glass systems.

## II. SIMULATIONS

The present simulations are carried out as state-of-the-art isothermal-isobaric $(N,p,T)$ calculations. The Newtonian equations of $N = 648$ atoms (130 Ni and 518 Zr) are numerically integrated by a fifth order predictor-corrector algorithm with time step $\Delta t = 2.5 \times 10^{-15}$ s in a cubic volume with periodic boundary conditions and variable box length $L$. With regard to the electron theoretical description of the interatomic potentials in transition metal alloys by Hausleitner and Hafner [29], we model the interatomic couplings as in [25] by a volume dependent electron-gas term $E_{vol}(V)$ and pair potentials $\phi(r)$ adapted to the equilibrium distance, depth, width, and zero of the Hausleitner-Hafner potentials [33] for $Ni_{20}Zr_{80}$ [1]. For this model, simulations were started through heating a starting configuration up to 2000 K which leads to a homogeneous liquid state. The system then is cooled continuously to various annealing temperatures with cooling rate $-\partial_t T = 1.5 \cdot 10^{12}$ K/s. Afterwards the obtained configurations at various annealing temperatures (here 1500-800 K) are relaxed by carrying out additional isothermal annealing runs. Finally the time evolution of these relaxed configurations is modelled and analyzed. More details of the simulations are given in [1].

## II. RESULTS AND DISCUSSIONS

From the simulated atomic motions in the computer experiments, the diffusion coefficients of the Ni and Zr species can be determined as the slope of the atomic mean square displacements in the asymptotic long-time limit

$$D_i(T) = \lim_{t \to \infty} \frac{(1/N_i)\sum_{\alpha=1}^{N_i} |r_\alpha(r) - r_\alpha(0)|^2}{6t}. \quad (1)$$

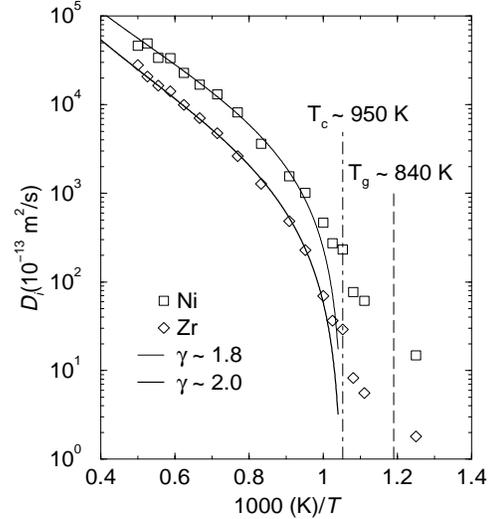

Fig.1 Diffusion Coefficients $D_i$ as a function of 1000/T. Symbols are MD results for Ni (square) and Zr (diamond); the full line are power-law approximations for Ni and for Zr resp.

Fig.1 shows the thus calculated diffusion coefficients of our $Ni_{20}Zr_{80}$ model for the temperature range between 800 and 2000 K. At temperatures above approximately 1250 K, the diffusion coefficients for both species run parallel to each other in the Arrhenius plot, indicating a fixed ratio $D_{Ni}/D_{Zr} \sim 2.5$ in this temperature regime. At lower temperatures, the Zr atoms have a lower mobility than the Ni atoms, yielding around 900 K a value of about 10 for $D_{Ni}/D_{Zr}$. That means, here the Ni atoms carry out a rather rapid motion within a relative immobile Zr matrix.

According to the MCT, above $Tc$ the diffusion coefficients follow a critical power law

$$D_i(T) \sim (T-Tc)^\gamma, \quad for \quad T > Tc \quad (2)$$

with non-universal exponent γ [30,31]. In order to estimate $Tc$ from this relationship, we have adapted the critical power law by a least mean squares fit to the simulated diffusion data for 1050 K and above. The results of the fit are included in Fig.1 by dashed lines. According to this fit, the system has a critical temperature of 950 K. The parameters γ turn out as 1.8 for the Ni subsystem and 2.0 for the Zr system.

Similar results for the temperature dependence of the diffusion coefficients have been found in MD simulations for other metallic glass forming systems, e.g., for $Ni_{50}Zr_{50}$ [26], for $Ni_xZr_{1-x}$ [3], $Cu_{33}Zr_{67}$ [32], or $Ni_{81}B_{19}$ [33]. In all cases, like here, a break is observed in the Arrhenius slope.

In the mentioned Zr-systems, this break is related to a change of the atomic dynamics around $Tc$ whereas for $Ni_{81}B_{19}$ system it is ascribed to $Tg$ (here $Tg$ denoted a caloric glass temperature) .As in [33] $Tc$ and $Tg$ apparently fall together, there is no serious conflict between the obervations.

A comment is necessary here concerning the comparison with experiments. Experimentally, around $Tg$ a break is observed in the Arrhenius slope [34-37] while the vicinity of the presumed $Tc$ not is covered by the present diffusion studies of the metallic glasses and melts. According to our understanding, a break in the Arrhenius slope is expected around $Tc$ as well as around $Tg$. As mentioned, the former reflects the change of the mechanism of the structural dynamics around $Tc$ described by the MCT. The latter is due to a change in frequency and phase space volume explored by the fluctuations when passing $Tg$. It depends significantly on the interplay between the time scale of the structural fluctuations and the hopping rate of the diffusing atoms, yielding that the break in the Arrhenius slope takes place at different temperatures for different diffusing species [34-36]. For the rather large cooling rates necessary in present MD simulations, it well may be that the corresponding high $Tg$ lies in a temperature region where significant changes in this interplay don't take place when passing the caloric $Tg$.

## IV. CONCLUSIONS

The present contribution reports results from MD simulations of a $Ni_{20}Zr_{80}$ computer model. The model is based on the electron theoretical description of the interatomic potentials for transition metal alloys by Hausleitner and Hafner [28]. There are no parameters in the model adapted to the experiments. Comparison of the calculated structure factors with experiments [22] indicates that the model is able to reproduce sufficiently well the steric relations and chemical order in the system, as far as visible in the structure factors.

We have found the $Tc$ values estimated from the dynamics in the undercooled melt when approaching $Tc$ from the high temperature side. The value is and $Tc \sim 950$ K from the diffusion coefficients. As discussed in [3], the $Tc$-estimates from the diffusion coefficients seem to depend on the upper limit of the temperature region taken into account in the fit procedure, where an increase in the upper limit increases the estimated $Tc$. Accordingly, there is evidence that the present value of 950 K may underestimate the true $Tc$ by about 10 to 50 K, as it based on an upper limit of 2000 K only. Taking this into account, the present estimates from the melt seem to lead to a $Tc$ value around 1000 K.

## REFERENCES

[1] A.B. Mutiara, Thesis, Universitaet Goettingen (2000).
[2] J. Jaeckle, Rep. Prog. Phys. 49, 171 (1986).
[3] U. Roessler and H. Teichler, Phys. Rev. B, 61, 394 (2000).
[4] W. Goetze, J. Phys. Cond. Mat. 11, A1 (1999).
[5] H.Z. Cummins, G. Li, Z.H. Hwang, G.Q.Shen, W.M. Du, J.Hernandes, and N.J. Tao, Z.Phys. B 103, 501 (1997).
[6] L.P. Kadanoff and J. Swift, Phys. Rev. 166, 89 (1968).
[7] K. Kawasaki, Phys. Rev. 150, 1 (1966); Ann. Phys. (N.Y.) \textbf{61}, 1 (1970).
[8] M.H. Ernst and J.R. Dorfman, J. Stat. Phys. 12, 311 (1975).
[9] U. Bengtzelius, W. Goetze, and A. Sjoelander, J. Phys. C 17, 5915 (1984).
[10] E. Leutheusser, Phys. Rev. A 29, 2765 (1984).
[11] W. Goetze and L. Sjoegren, Rep. Prog. Phys. 55, 241 (1992)
[12] P.S. Das and G.F. Mazenko, Phys. Rev. A 34, 2265 (1986).
[13] L. Sjoegren, Z. Phys. B 79, 5 (1990).
[14] T. Gleim and W. Kob, Eur. Phys. J. B 13, 83 (2000).
[15] A. Meyer, R. Busch, and H. Schober, Phys. Rev. Lett. 83, 5027 (1999); A. Meyer, J. Wuttke, W. Petry, O.G. Randl, and H. Schober, Phys. Rev. Lett. 80, 4454 (1998).
[16] H.Z. Cummins, J. Phys.Cond.Mat. 11, A95 (1999).
[17] H. Teichler, Phys. Rev. Lett. 76, 62(1996).
[18] H. Teichler, Phys. Rev. E 53, 4287 (1996).
[19] J.L. Barrat and A. Latz, J. Phys. Cond. Matt. 2, 4289 (1990).
[20] M. Fuchs, Thesis, TU-Muenchen (1993); M. Fuchs and A. Latz, Physica A 201, 1 (1993).
[21] M. Nauroth and W. Kob, Phys. Rev. E 55, 657 (1997).
[22] M. Kuschke, Thesis, Universitaet Stuttgart (1991).
[23] Yan Yu, W.B. Muir and Z. Altounian, Phys. Rev. B 50, 9098 (1994).
[24] B. Boeddeker, Thesis, Universitaet Goettingen (1999); B. Boeddeker and H. Teichler, Phys. Rev. E 59, 1948 (1999)
[25] H. Teichler, phys. stat. sol. (b) 172, 325 (1992)
[26] H. Teichler, in: Defect and Diffusion Forum 143-147, 717 (1997)
[27] H. Teichler, in: Simulationstechniken in der Materialwissenschaft, ed. by P. Klimanek and M. Seefeldt (TU Bergakademie, Freiberg, 1999).
[28] H. Teichler, Phys. Rev. B 59, 8473 (1999).
[29] Ch. Hausleitner and Hafner, Phys. Rev. B 45, 128 (1992).
[30] J.-P. Hansen and S. Yip, Transp. Theory Stat. Phys. 24, 1149 (1995).
[31] W. Kob and H.C. Andersen, Phys. Rev. E 51, 4626 (1995).
[32] C. Gaukel, Thesis, TU-Aachen (1998).
[33] L.D. van Ee, Thesis, TU-Delft (1998).
[34] K. Knorr, Thesis, Universitaet Muenster (2000).
[35] U. Geyer, S. Schneider, W. L. Johnson, Y. Qiu, T. A. Tombrello, and M.-P. Macht, Phys. Rev. Lett. 75, 2364 (1995).
[36] P. Filietz, M. Macht, V. Naundorf and G. Frohberg, J. Non-Cryst. Solids 250, 674 (1999).
[37] X.-P. Tang, U. Geyer, R. Busch, W.L. Johnson and Yue Wu, Nature 402, 160 (1999).